%% file: main.tex
\documentclass[runningheads]{llncs}
\usepackage{graphicx}
\usepackage{amsmath,amssymb,amsfonts}
\usepackage{appendix}
\usepackage{comment}
\usepackage{enumitem}
\usepackage{tablefootnote}
\usepackage{booktabs,multirow,multicol}
\usepackage[colorlinks=true, allcolors=blue]{hyperref}
\usepackage{caption}
\usepackage{cleveref}
\usepackage{minted}

\setminted{
fontsize=\footnotesize,
breaklines,
bgcolor=lightgray,
framesep=2mm
}
\usepackage{./common-unicode}

\newcommand{\rad}{\mathrm{rad}}

\begin{document}
\title{Formalizing Mason--Stothers Theorem and its Corollaries in Lean 4}
\titlerunning{Formalizing Mason--Stothers Theorem and its Corollaries in Lean 4}
\author{Jineon Baek\inst{1}\orcidID{0000-0002-5799-4902} \and
Seewoo Lee\inst{2}\orcidID{0000-0002-5710-2257}}
\authorrunning{J. Baek and S. Lee}
\institute{
Yonsei University
 \email{jineon@yonsei.ac.kr} \and
University of California, Berkeley \email{seewoo5@berkeley.edu}
}
\maketitle              % typeset the header of the contribution
\begin{abstract}

The \emph{ABC conjecture} implies many conjectures and theorems in number theory, including the celebrated Fermat's Last Theorem.
The \emph{Mason--Stothers Theorem} is a function field analogue of the ABC conjecture that admits a much more elementary proof with many interesting consequences,
including a polynomial version of Fermat's Last Theorem.
While years of dedicated effort are expected for a full formalization of  
Fermat's Last Theorem, the simple proof of the Mason--Stothers Theorem and its corollaries calls for an immediate formalization.

We formalize an elementary proof by Snyder in Lean 4, and also formalize many consequences of Mason--Stothers, including nonsolvability of Fermat--Cartan equations in polynomials, nonparametrizability of a certain elliptic curve, and Davenport's Theorem.
We compare our work to existing formalizations of the Mason--Stothers by Eberl in Isabelle and Wagemaker in Lean 3 respectively.
% Our formalization is based on the \texttt{mathlib4} library of Lean 4, and is currently being ported back to \texttt{mathlib4}.
Our formalization has been integrated into the \texttt{mathlib} library of Lean 4.

\end{abstract}

\input{1introduction}

\input{2statement-informal}

\input{3mason-stothers-proof-informal}
\input{4prelims}

\input{5mason-stothers-formal}

\input{6corollaries}
\input{7comparison}
\input{8futureworks}

\subsection*{Acknowledgement}
We thank Kevin Buzzard for suggesting the project.
Also, we thank Thomas Browning for his help in simplifying the formalization of  Davenport's theorem.
We also thank the reviewers of \texttt{mathlib} who helped improving our codes and porting them into \texttt{mathlib}, including Riccardo Brasca, Johan Commelin, Yaël Dillies, Diamiano Testa, Ruben Van de Velde, Eric Weiser, Junyan Xu, and Andrew Yang.
Jineon Baek acknowledges the support from Korea Foundation for Advanced Studies during the completion of this work.

\input{appendix}

\bibliographystyle{splncs04}
\bibliography{ref}

\end{document}

%% file: 1introduction.tex
\section{Introduction}\label{sec:1}

% \textcolor{red}{Things to be updated in the paper:
% \begin{enumerate}
%     \item Add/update comparison with previous formalizations
%     \item Add/update explanations on the FLT formalization (Buzzard)
%     \item Refactor paper structure
% \end{enumerate}
% }

In 1985, Oesterlé and Masser proposed the \emph{ABC conjecture} \cite{masser1985open,oesterle1988nouvelles}:
\begin{conjecture}[ABC conjecture]
For every positive real number $\varepsilon > 0$, there exist only finitely many triples of coprime integers $(a, b, c)$ such that $a+b = c$ and 
\[
    c  > \rad(abc)^{1 + \varepsilon}.
\]
Here, $\rad(n) = \prod_{p|n} p$ is the product of all prime factors of $n$.
\end{conjecture}
The conjecture implies many deep theorems or conjectures in number theory.
For example, Fermat's Last Theorem (FLT) for exponent $n \geq 6$ is a direct corollary of an explicit quantitative version of the ABC conjecture \cite{granville02}, while the currently known proof by Wiles \cite{wiles95} and Taylor--Wiles \cite{taylor95} requires heavy machinery (See \cite{cornell13} for a detailed survey).
Also, Roth's theorem \cite{roth55} and Faltings' theorem \cite{faltings83} both follow from the ABC conjecture \cite{frankenhuysen99}; note that the proof of each theorem earned its corresponding author a Fields medal.

% In essence, the conjecture says that the sum of two coprime integers, both divisible by high power of primes, is unlikely to be also divisible by high powers of primes.
% Despite its elementary nature, the conjecture remains wide open with only exponential bounds established so far \cite{stewart}.

In number theory, there is a strong analogy between (finite extensions of) $\mathbb{Q}$ and (finite extensions of) a rational function field $k(t)$ over a field $k$.
Under this analogy, profound statements on integers $\mathbb{Z}$, such as the Riemann Hypothesis, the Birch and Swinnerton-Dyer conjecture, or the Langlands program, have analogous statements \cite{deligne74,deligne80,lafforgue02,tate95} on the integral ring $k[t]$ of the rational function field $k(t)$.
The analogs of such conjectures often turn out to be true and easier to prove in general.

In this line, Stothers proved the polynomial analog of the ABC conjecture in 1981 \cite{stothers81}, and Mason rediscovered it in a more general form in 1983  \cite{mason84}, even before Osterl\'e and Masser proposed the ABC conjecture.

\begin{definition}
\label{def:radical}
Let $k$ be any field. For any nonzero $f \in k[X]$,  define the radical $\rad (f)$ of $f$ as the product of all irreducible monic factors of $f$ not counting multiplicity.
\end{definition}
\begin{theorem}[Mason--Stothers]
\label{thm:Mason--Stothers}
Let $k$ be any field and $a, b, c \in k[t]$ be non-zero, pairwise coprime polynomials satisfying $a + b + c = 0$.
Then we either have $a' = b' = c' = 0$ where $f'$ denotes the (formal) derivative of $f \in k[t]$ by $t$, or
\[
\max\{\deg(a), \deg(b), \deg(c)\} < \deg (\rad(abc)).
\]
\end{theorem}

Mason and Stothers proved the theorem using algebro-geometric methods, and subsequently Snyder discovered a short and purely elementary proof \cite{snyder00}.
Like the ABC conjecture, the Mason--Stothers theorem has a lot of interesting consequences, including the following:
\begin{enumerate}
    \item A polynomial version of Fermat's Last Theorem. More generally, the non-solvability of the Fermat--Catalan equation $ua^p + vb^q + wc^r = 0$ over $a, b, c \in k[t]$ with nonzero constants $u, v, w \in k$ and powers $p, q, r \in \mathbb{N}$ satisfying $1 / p + 1 / q + 1 / r \leq 1$ (see Theorem \ref{cor:fermat-catalan}).
    \item Non-parametrizability of the elliptic curve $y^2 = x^3 + 1$ by rational functions $x, y \in k(t)$ (see Theorem \ref{cor:elliptic}).
    \item Davenport's theorem, initially conjectured by Birch et al \cite{birch65},  that for any polynomials $f, g \in \mathbb{C}[t]$ we have $\deg{(f^3 - g^2)} \geq \frac{1}{2}\deg{f} + 1$ (see Theorem \ref{cor:davenport}).
\end{enumerate}

We give a fully documented Lean 4 formalization of the Mason--Stothers theorem on fields of arbitrary characteristic.
Also, we formalize the aforementioned corollaries of the theorem to demonstrate the power of the Mason--Stothers theorem (see Theorem \ref{cor:fermat-catalan}, \ref{cor:elliptic}, and \ref{cor:davenport}).
The original code is hosted in
\begin{center}
\url{https://github.com/seewoo5/lean-poly-abc}
\end{center}
% and is currently being ported to \texttt{mathlib} (see \Cref{sec:porting}).
and the proofs of the Mason--Stothers theorem and polynomial FLT are fully integrated into \texttt{mathlib} (see \Cref{sec:porting}).

The Mason--Stothers theorem was already formalized by Eberl in Isabelle \cite{eberl17} and Wagemaker in Lean 3 \cite{wagemaker18}.
We give a detailed comparison between our work and theirs in \Cref{sec:comparison}.
In short, our formalization works for arbitrary characteristic (see \Cref{subsec:comp-wagemaker}), is compatible with the \texttt{mathlib} library of Lean 4, includes variants of the Mason--Stothers (\Cref{thm:mason-stothers-nocoprime}). We also formalize several interesting corollaries including the polynomial FLT with a slightly stronger conclusion than the existing formalization (see \Cref{subsec:comp-eberl}).

%% file: 2statement-informal.tex
\section{Statements of the Theorem and its Corollaries}
\label{sec:statements}

The precise statement of Mason--Stothers theorem we formalize is \Cref{thm:Mason--Stothers}, which holds for arbitrary field $k$.
Note that most literature either assumes that $k$ is of characteristic zero or is algebraically closed \cite{stothers81,snyder00}.

If $k$ has characteristic zero, the condition $a' = b' = c' = 0$ in \Cref{thm:Mason--Stothers} is equivalent to $a, b, c$ being constants.
If $k$ has characteristic $p > 0$, then the condition $f' = 0$ for $f = a, b, c$ is equivalent to $f(t) = f_0(t^p)$ being a polynomial of $t^p$.
Indeed,
$$
(a, b, c) = (-1, -x^p, 1 + x^p) = (-1, -x^p, (1 + x)^p)
$$ is an example satisfying $a + b + c = 0$ and $a' = b' = c' = 0$, but
$$
\max\{\deg(a), \deg(b), \deg(c)\} + 1 = p + 1 > 2 = \deg (\rad(abc)).
$$

We now state the corollaries of the Mason--Stothers theorem mentioned in \Cref{sec:1} precisely.
Their proofs can be found in \Cref{sec:formal-corr}.

The \emph{Fermat--Catalan conjecture} is a generalization of Fermat's Last Theorem stating that the equation $a^p + b^q = c^r$ has only finitely many solutions $(a, b, c, p, q, r)$ in positive integers satisfying $1/p + 1/q + 1/r < 1$ \cite{granville02}. 
The following \Cref{cor:fermat-catalan} is a polynomial variant which is known to be true.

\begin{theorem}[Fermat--Catalan Conjecture for Polynomials]
\label{cor:fermat-catalan}
Let $k$ be any field.
Let $p, q, r \geq 1$ be integers not divisible by the characteristic of $k$
such that $1/p + 1/q + 1/r \leq 1$.
Let $u, v, w \in k$ be arbitrary nonzero constants.
Then any triple $(a, b, c)$ of nonzero and pairwise coprime polynomials in $k[t]$ satisfying $ua^p + vb^q + wc^r = 0$ consists of constants $a, b, c \in k$.
\end{theorem}

Let $u = v = -w = 1$ and $p = q = r = n \geq 3$ in \Cref{cor:fermat-catalan} to recover the Fermat's Last Theorem for polynomials.

\begin{corollary}[Fermat's Last Theorem for Polynomials]
\label{cor:fermat}
Let $k$ be any field.
Let $n \geq 3$ be any integer not divisible by the characteristic of $k$.
Then any triple $(a, b, c)$ of nonzero and pairwise coprime polynomials in $k[t]$ satisfying $a^n + b^n = c^n$ should be constants $a, b, c \in k$.
\end{corollary}

Using Theorem \ref{cor:fermat-catalan}, we can obtain the following corollary.
\begin{theorem}[Non-parametrizablility of an Elliptic Curve]\label{cor:elliptic}
Let $k$ be a field of characteristic $\neq 2, 3$.
If rational functions $f(t), g(t) \in k(t)$ satisfy $g(t)^2 = f(t)^3 + 1$, then both $f(t)$ and $g(t)$ are constants in $k$.
\end{theorem}

In other words, the elliptic curve defined by the Weierstrass equation $y^2 = x^3 + 1$ is not parametrizable by non-constant rational functions in $k(t)$.

Another interesting corollary of the Mason--Stothers theorem is the following theorem by Davenport \cite{davenport65}, initially conjectured \cite{birch65} by Birch et al.
This theorem motivated Stothers' proof of the Mason--Stothers theorem \cite{stothers81}.

\begin{theorem}[Davenport]\label{cor:davenport}
% \footnote{The statement is slightly different from the original theorem in \cite{stothers81}. Stothers proved the theorem for pairs of polynomials that are not necessarily coprime, but he assumed that the field $k$ has characteristic zero. In our proof, we extended the theorem to hold for positive characteristic as well, but we included the hypothesis that $f$ and $g$ are coprime. It is possible to remove this hypothesis when the characteristic is zero by proving a slightly different version of the Mason--Stothers theorem, but we have not included this in our formalization.}
Let $k$ be a field of characteristic zero. Let $f(t), g(t) \in k[t]$ be non-constant polynomials such that $f^3 \neq g^2$.
Then 
\[
    \deg(f^3 - g^2) \geq \frac{1}{2} \deg(f) + 1.
\]
\end{theorem}

% In this paper, we describe our formalization of Snyder's proof of the Mason--Stothers theorem (Theorem \ref{thm:Mason--Stothers}) and its corollaries (Theorem \ref{cor:fermat-catalan}, \ref{cor:elliptic}, and \ref{cor:davenport}) including the analog of FLT over polynomial rings. 
% The formalization is done in Lean 4 using the \texttt{mathlib} mathematics library  \cite{mathlib20}.
% The code is hosted in \url{https://github.com/seewoo5/lean-poly-abc}.

%% file: 3mason-stothers-proof-informal.tex
\section{Mathematical Proof of the Mason--Stothers Theorem}
\label{sec:pf-mason-stothers}

We summarize the proof of the Mason--Stothers theorem (Theorem \ref{thm:Mason--Stothers}) in Lemmermeyer's note \cite{franz} that we formalize.
\begin{definition}
\label{def:wronskian}
The \emph{Wronskian} of two polynomials $a, b \in k[t]$ is $W(a, b) = ab' - a'b$.
\end{definition}
From $a + b + c = 0$, we can check that the values $W(a, b)$, $W(b, c)$, and $W(c, a)$ are all equal. Denote the common value as $W$.

Next, we observe the following property.
\begin{lemma}
\label{lem:main}
For a nonzero polynomial $a \in k[t]$, $a /\rad(a)$ divides $a'$.
\end{lemma}
\begin{proof}
We can use the prime factorization of $a$.
Let $a = u p_1 ^{e_1}  p_2 ^{e_2}  \cdots  p_m ^{e_m}$ be a factorization with unit $u$ and primes $p_i \in k[x]$ of exponents $e_i > 0$.
Then the product rule of derivative gives $a' = \sum_{i=1}^m u e_i p_i'  p_1 ^{e_1}  p_2 ^{e_2} \cdots p_i ^{e_i - 1} \cdots  p_m ^{e_m}$
which is divisible by $a /\rad(a) = p_1 ^{e_1 - 1}  p_2 ^{e_2 - 1} \cdots p_m ^{e_m - 1}$.
\end{proof}
% For a proof, we can use the prime factorization of $a$.
% Let $a = u p_1 ^{e_1}  p_2 ^{e_2}  \cdots  p_m ^{e_m}$ be a factorization with unit $u$ and primes $p_i \in k[x]$ of exponents $e_i > 0$.
% Then the product rule of derivative gives $a' = \sum_{i=1}^m u e_i p_i'  p_1 ^{e_1}  p_2 ^{e_2} \cdots p_i ^{e_i - 1} \cdots  p_m ^{e_m}$
% which is divisible by $a /\rad(a) = p_1 ^{e_1 - 1}  p_2 ^{e_2 - 1} \cdots p_m ^{e_m - 1}$.
Since $a / \rad(a)$ divides both $a$ and $a'$, an immediate corollary is that
\begin{lemma}
For any nonzero $a \in k[t]$, $a /\rad(a)$ divides $W(a, b) = ab' - a'b$.
\end{lemma}
% because $a /\rad(a)$ divides both $a$ and $a'$.

\begin{proof}[Proof of \Cref{thm:Mason--Stothers}]
The pairwise coprime polynomials $a / \rad(a)$, $b / \rad(b)$, and $c / \rad(c)$ all divide $W$, so their product $abc / \rad(abc)$ should also divide $W$. This is the key step of the proof.

We split the proof into two cases, depending on whether $W$ is zero or not.
% Divide the case into whether $W$ is zero or not.
If $W = 0$, then $W(a, b) = 0$ implies $a b' = a' b$, and since $a$ and $b$ are coprime $a$ divides $a'$ and so $a' = 0$. Likewise, from $W = 0$ we also get $b' = c' = 0$.

Now assume $W \neq 0$. Then $abc / \rad(abc)$ dividing $W$ implies
\begin{align*}
    &\deg(a) + \deg(b) + \deg(c) - \deg(\rad(abc)) = \deg \left(\frac{abc}{\rad(abc)}\right) \\
    &\le \deg W = \deg W(a, b) < \deg(a) + \deg(b).
\end{align*}
The first inequality follows from divisibility, and the second inequality follows from the definition of Wronskian and $a \neq 0$.
Hence we have $\deg(c) < \deg \rad(abc)$. The same argument with $W = W(b, c)$ and $W = W(c, a)$ gives
\[
\max\{\deg(a), \deg(b), \deg(c)\} + 1\le \deg(\rad(abc)).
\]
\end{proof}

%% file: 4prelims.tex
\section{Basic Definitions}
\label{sec:prelim-formalize}

In this section, we explain formalizations of the basic definitions and lemmas for the proof of \Cref{thm:Mason--Stothers}.
The first step is to develop an interface for the radical (\Cref{def:radical}) and Wronskian (\Cref{def:wronskian}) of polynomials.

\subsection{Wronskian}
\label{subsec:wronskian}

We formalize the Wronskian $W(a, b)$ of any two polynomials $a, b \in R[X]$ with coefficients in an arbitrary commutative ring $R$.

\begin{minted}{lean}
variable {R : Type*} [CommRing R]

def wronskian (a b : R[X]) : R[X] :=
  a * (derivative b) - (derivative a) * b
\end{minted}

We show that the degree of Wronskian $W(a, b)$ is strictly smaller than $\deg(a) + \deg(b)$, which was one of the last steps in our proof of \Cref{thm:Mason--Stothers}.

\begin{minted}{lean}
theorem wronskian.natDegree_lt_add {a b : R[X]}
    (hw : wronskian a b ≠ 0) :
  natDegree (wronskian a b) < natDegree a + natDegree b
\end{minted}

For a polynomial \texttt{a} in Lean 4's \texttt{mathlib}, both \texttt{degree a} and \texttt{natDegree a} denote the degree of \texttt{a}.
The difference between the two is that the \texttt{natDegree} has type $\mathbb{N}$ of natural numbers, while the \texttt{degree} has type \texttt{WithBot $\mathbb{N}$} which is $\mathbb{N}$ equipped with $-\infty$.
The \texttt{natDegree} of zero polynomial is defined as 0, while the \texttt{degree} of that is defined as $-\infty$.
While \texttt{degree} is mathematically more natural, we opt to use \texttt{natDegree} as its type $\mathbb{N}$ is much easier to work in Lean 4 than the extended type \texttt{WithBot $\mathbb{N}$}.

We use that $W(a, b) = W(b, c)$ for any $a, b, c \in R[X]$ with $a + b + c = 0$ in our proof.
This identity actually holds for any alternating bilinear map $B: M \times M \to R$ on any $R$-module $M$.
Thus we add the general theorem in the relevant file of \texttt{mathlib}.\footnote{\texttt{Mathlib.LinearAlgebra.BilinearForm.Properties}}

\begin{minted}{lean}
theorem eq_of_add_add_eq_zero [IsCancelAdd R] {a b c : M} (H : B.IsAlt) (hAdd : a + b + c = 0) : B a b = B b c
\end{minted}

Note that the above theorem is stated for a slightly general class of $R$ called \texttt{IsCancelAdd}, where the additive structure $(R, +)$ is not necessarily a group but still satisfy the cancellation law: for any $x, y, z \in R$, $x + z = y + z \Rightarrow x = y$.

\subsection{Radical}
\label{subsec:radical}

Recall that for any field $k$ and nonzero $f \in k[X]$, its radical $\rad (f)$ is defined as the product of all irreducible monic factors of $f$ not counting multiplicity. In fact, such a definition works over any multiplicative monoid $M$ with zero that is

\begin{enumerate}
    \item commutative,
    \item cancellative ($ab = ac \Rightarrow b=c$ for nonzero $a$),
    \item a \emph{unique factorization monoid}, meaning that each nonzero element admits a unique factorization into irreducible elements, and
    \item 
    % a \emph{normalization monoid}, 
    equipped with a map $u : M \setminus \{0\} \rightarrow M^*$ to the set of units $M^*$ of $M$ which preserves multiplication. The map $x \mapsto u(x)^{-1} x$ is then called the \emph{normalization map}.
\end{enumerate}
In particular, if $M$ is the polynomial ring $k[X]$, the map $u : k[X] \setminus \{0\} \rightarrow k^\times$ in (4) reads the leading coefficient of a nonzero polynomial. The corresponding normalization map $a \mapsto u(a)^{-1} a$ sends a polynomial $a$ to its unique monic scalar multiple.

In \texttt{mathlib}, these assumptions can be imposed on a monoid $M$ by using the following instances on $M$.
\begin{enumerate}[label=(\roman*)]
    \item \texttt{CancelCommMonoidWithZero} (for 1 and 2)
    \item \texttt{UniqueFactorizationMonoid} (for 3)
    \item \texttt{NormalizationMonoid} (for 4)
\end{enumerate}

To define the radical of $a \in M$, we first extract the multiset of 
normalized factors of $a$ from \texttt{mathlib} using \texttt{normalizedFactors a}.
% Here, a multiset is essentially a set allowing duplicated elements.
Note that multiset in \texttt{mathlib} is finite, and we can convert it to the corresponding finite set using \texttt{.toFinset} to get rid of duplicated elements.
Then we multiply all the elements in the finite set to get the radical of $a \in M$.
% Then we convert it to a finite set (\texttt{.toFinset}) to get rid of duplicated elements, 
% and multiply them all to get the radical of $a \in M$.

\begin{minted}{lean}
variable {M : Type*} [CancelCommMonoidWithZero M] [UniqueFactorizationMonoid M] [NormalizationMonoid M]

/-- Prime factors of `a` are monic factors of `a` without duplication. -/
def primeFactors (a : M) : Finset M :=
  (normalizedFactors a).toFinset

/-- Radical of `a` is a product of prime factors of `a`. -/
def radical (a : M) : M :=
  (primeFactors a).prod id
\end{minted}

When $M$ equals the polynomial ring $k[X]$, any radical is a monic polynomial, and $\text{rad}(c) = 1$ for any constant $c \in k$ including zero.\footnote{The set of normalized factors of zero or a unit is an empty set in \texttt{mathlib}. The product of elements in an empty set is then defined as 1 in \texttt{mathlib}.}

Radical satisfies the power law $\text{rad} (a^n) = \text{rad} (a)$ for $n \geq 1$.
\begin{minted}{lean}
theorem radical_pow (a : M) {n : Nat} (hn : 0 < n) :
    radical (a ^ n) = radical a
\end{minted}

Also, $\text{rad}(a)$ divides $a$.
Although this is obvious from unique factorization, Lean is not aware of this intuition.
A formal proof uses basic lemmas in \texttt{mathlib}\footnote{e.g., If a multiset $A$ is contained in $B$, the product of elements in $A$ divides that of $B$ (\texttt{Multiset.prod\_dvd\_prod\_of\_le}).} to boil down the proof to that the \texttt{Multiset} $S$ of prime factors of $a$ contains,
as a subset, the same set $S$ with duplicated elements removed.
\begin{minted}{lean}
theorem radical_dvd_self (a : M) : radical a | a
\end{minted}

Once we restrict our attention to a \emph{commutative domain} $R$ with unique factorization, we can also prove multiplicativity of the radical for coprime elements $a, b \in R$, i.e. $\text{rad} (ab) = \text{rad} (a) \text{rad} (b)$.
We also have $\rad (-a) = \rad(a)$. These basic lemmas will be used frequently in the main proof.

\begin{minted}{lean}
variable {R : Type*} [CommRing R] [IsDomain R] [NormalizationMonoid R] [UniqueFactorizationMonoid R]

theorem radical_mul {a b : R} (hc : IsCoprime a b) :
  radical (a * b) = (radical a) * (radical b)

theorem radical_neg {a : R} : radical (-a) = radical a
\end{minted}

The following seems obvious to the human eye.
\begin{lemma}
\label{lem:nzdeg}
For any field $k$ and a polynomial $a \in k[X]$ of degree $\geq 1$, the degree of its radical $\rad(a)$ is also at least one.
\end{lemma}
A formal proof of \Cref{lem:nzdeg} requires more work, however. We need to explicitly take a prime factor $p$ of $a$ and show that it is also a prime factor of $\rad(a)$.
We first show that for any element $a$ of a general monoid $M$, a prime $p$ divides $a$ if and only if it divides $\rad(a)$; this is done by using that the prime divisors of $a$ and $\rad(a)$ are the same.
\begin{minted}{lean}
theorem prime_dvd_radical_iff {a p : M} (ha : a ≠ 0) (hp : Prime p) :
    p | radical a ↔ p | a
\end{minted}
We then use it to show that a nonzero $a \in M$ is a unit if and only if $\rad(a)$ is. Note that an element of $M$ is a unit if and only if it has no prime divisors.
\begin{minted}{lean}
theorem radical_isUnit_iff {a : M} (h : a ≠ 0) :
    IsUnit (radical a) ↔ IsUnit a
\end{minted}
Then we specialize it to $M = k[X]$ and use that a nonzero $a \in k[X]$ is a unit if and only if its degree is zero, proving \Cref{lem:nzdeg}.
\begin{minted}{lean}
lemma natDegree_radical_eq_zero_iff {a : k[X]} :
    (radical a).natDegree = 0 ↔ a.natDegree = 0
\end{minted}

The fraction $f / \rad (f)$ is a polynomial which will be used frequently in the proof. We define this as \texttt{divRadical f} in our formalization.

\begin{minted}{lean}
def divRadical (a : k[X]) : k[X] := a / radical a
\end{minted}
The division notation actually denotes the quotient of two polynomials as in the Euclidean division algorithm. 
Using that the radical divides the polynomial (\texttt{radical\_dvd\_self}), we prove lemmas that introduces and eliminates \texttt{divRadical f} as it is multiplied by \texttt{radical f}.
With this, we do not need to work with division explicitly and only work with multiplications, which are easier to handle with Lean 4 tactics like \texttt{ring}.

\begin{minted}{lean}
theorem eq_divRadical {a x : k[X]} (h : (radical a) * x = a) : x = divRadical a

theorem mul_radical_divRadical (a : k[X]) :
    (radical a) * (divRadical a) = a
\end{minted}

Now we need to prove \Cref{lem:main} that for any $a \in k[X]$, $a / \rad(a)$ divides $a'$.
\begin{minted}{lean}
theorem divRadical_dvd_derivative (a : k[X]) : 
    (divRadical a) | (derivative a)
\end{minted}
Our formalization does not explicitly use the factorization $a = u p_1 ^{e_1}  p_2 ^{e_2}  \cdots  p_m ^{e_m}$ which is somewhat cumbersome to work with in Lean.
Instead, we use the \emph{coprime induction} in \texttt{mathlib}.\footnote{Available as \texttt{induction\_on\_coprime} in {\texttt{mathlib}}.}
We first prove the result for units $a = u$ and prime powers $a = p^e$.
Then we show that for any coprime $a, b$ satisfying the lemma, their product $ab$ also satisfies the lemma.
This makes the derivative $(ab)' = a'b + ab'$ much easier to manipulate than the derivative of the full factorization.

%% file: 5mason-stothers-formal.tex
\section{Formalization of the Mason--Stothers theorem}
\label{sec:mason-stothers-formal}

Finally, Mason--Stothers theorem is formalized as follows.
Note that \texttt{k} denotes any field of arbitrary characteristic \texttt{[Field k]} assumed.

\begin{minted}{lean}
variable {k : Type*} [Field k]

theorem Polynomial.abc {a b c : k[X]}
  (ha : a ≠ 0) (hb : b ≠ 0) (hc : c ≠ 0) 
  (hab : IsCoprime a b) (hsum : a + b + c = 0) :
  (derivative a = 0 ∧ derivative b = 0 ∧ derivative c = 0) ∨
    Nat.max₃ a.natDegree b.natDegree c.natDegree + 1 ≤ (radical (a * b * c)).natDegree
\end{minted}

We only require coprimality of $a$ and $b$, as $\gcd(b, c) = \gcd(c, a) = 1$ can be deduced from $\gcd(a, b)$ and $a + b + c = 0$.
Because $a, b, c$ are nonzero, there is no mathematical difference in using \texttt{natDegree} instead of \texttt{degree}.
  
To formalize Mason--Stothers, we first formalize the proof of $abc/\rad(abc) | W$ mentioned as the key step of the proof in \Cref{sec:pf-mason-stothers}.
Then we prove an auxiliary lemma below that derives $\deg(c) < \deg(\rad(abc))$ from $abc/\rad(abc) | W$.

\begin{minted}{lean}
private theorem abc_subcall
  {a b c w : k[X]} {hw : w ≠ 0} (wab : w = wronskian a b)
  (ha : a ≠ 0) (hb : b ≠ 0) (hc : c ≠ 0)
  (hab : IsCoprime a b) (hbc : IsCoprime b c) (hca : IsCoprime c a)
  (abc_dr_dvd_w : (a * b * c).divRadical | w) :
    c.natDegree + 1 ≤ (radical (a * b * c)).natDegree
\end{minted}

Once the auxiliary lemma \texttt{abc\_subcall} is shown, we apply this three times to the permuted triples $(a, b, c)$, $(b, c, a)$, and $(c, a, b)$ to prove the full Mason--Stothers.
While it is evident that the conditions of \texttt{abc\_subcall} are symmetric,
we have to manually permute them in our formalization (e.g., change the product \texttt{a * b * c} to \texttt{b * c * a} in \texttt{abc\_dr\_dvd\_w}).
This, however, costs much less than repeating the whole argument three times.

%% file: 6corollaries.tex
\section{Formalization of Corollaries}
\label{sec:formal-corr}

We also formalize multiple corollaries of Mason--Stothers (\Cref{cor:fermat-catalan,cor:elliptic,cor:davenport}).

\input{61fermat-catalan}
\input{62elliptic}

\input{63davenport}

%% file: 61fermat-catalan.tex
\subsection{Fermat--Catalan Conjecture for Polynomials (\Cref{cor:fermat-catalan})}
\label{subsec:pf-fermat-catalan}

\subsubsection{Mathematical Proof}
\label{subsubsec:pf-fermat-catalan}

\Cref{cor:fermat-catalan} basically follows from Mason--Stothers applied to the triple $(ua^p, vb^q, wc^r)$.
Let 
$$m = \max \{\deg (ua^p), \deg(vb^q), \deg(wc^r)\} = \max \{ p\deg (a), q \deg (b), r \deg (c)\}.$$
If the inequality $m < \deg (\rad(a^pb^qc^r))$ holds, then we have
\begin{align*}
     m &<\deg (\rad(a^pb^qc^r)) = \deg(\rad(abc)) \leq \deg(abc) \\
     &= \deg(a) + \deg (b) + \deg(c) = \frac{1}{p} \cdot p\deg(a) + \frac{1}{q} \cdot q\deg(b) + \frac{1}{r} \cdot r\deg(c) \\
     &\leq \left(\frac{1}{p} + \frac{1}{q} + \frac{1}{r} \right)m
\end{align*}
which is a contradiction. So by Mason--Stothers it follows that $(a^p)' = (b^q)' = (c^r)' = 0$. As none of $p, q,$ or $r$ are zero in $k$, we conclude $a' = b' = c' = 0$. If the characteristic of $k$ is zero, then $a' = b' = c' = 0$ immediately implies that $a, b, c$ are constants.

If the characteristic $\ell$ of $k$ is positive, we need an extra infinite descent argument to show that $a, b, c$ are constants.
For $f = a, b, c$, that $f' = 0$ in $k[t]$ implies the existence of $f_1 \in k[t]$ such that $f(t) = f_1(t^\ell)$.
Hence we have $ua_1(t^\ell)^p + v b_1(t^\ell)^q + w c_1(t^\ell)^r = 0$.
Substitution $T = t^\ell$ gives $ua_1(T)^p + vb_1(T)^q + w c_1(T)^r = 0$, giving rise to a new nontrivial solution $(a_1, b_1, c_1)$ with strictly smaller yet nonzero degrees.
Repeated application of this descent in degree leads to contradiction.

\subsubsection{Formalization}

The full statement of \Cref{cor:fermat-catalan} we formalize is the following.

\begin{minted}{lean}
theorem Polynomial.flt_catalan
  {p q r : ℕ} (hp : 0 < p) (hq : 0 < q) (hr : 0 < r)
  (hineq : q * r + r * p + p * q ≤ p * q * r)
  (chp : ¬ringChar k | p) (chq : ¬ringChar k | q) (chr : ¬ringChar k | r)
  {a b c : k[X]} (ha : a ≠ 0) (hb : b ≠ 0) (hc : c ≠ 0) (hab : IsCoprime a b)
  {u v w : k} (hu : u ≠ 0) (hv : v ≠ 0) (hw : w ≠ 0)
  (heq : C u * a ^ p + C v * b ^ q + C w * c ^ r = 0) :
  a.natDegree = 0 ∧ b.natDegree = 0 ∧ c.natDegree = 0
\end{minted}

We state the inequality $1 / p + 1/q + 1/r \leq 1$ 
as $qr + rs + sp \leq pqr$ instead, as this is expressible purely in integers which is easier to work with in Lean 4.
In \texttt{mathlib}, for any element \texttt{u : k} in field $k$ the notation \texttt{C u : k[X]} denotes the corresponding constant polynomial in the ring.

To formalize \Cref{cor:fermat-catalan}, we prove that, under the hypothesis of \Cref{cor:fermat-catalan}, $a' = b' = c' = 0$.
\begin{minted}{lean}
theorem Polynomial.flt_catalan_deriv
  /-...same condition as flt_catalan...-/ :
    derivative a = 0 ∧ derivative b = 0 ∧ derivative c = 0
\end{minted}

We then formalize the infinite descent argument above to show that the degree of $a$ is zero.
If the characteristic of $k$ is nonzero, we apply a strong induction\footnote{\texttt{Nat.case\_strong\_induction\_on} in \texttt{mathlib}} on the degree of $a$.
\begin{minted}{lean}
theorem Polynomial.flt_catalan_aux
  /-...same condition as flt_catalan...-/ :
    a.natDegree = 0
\end{minted}
Then we use this auxiliary step three times to formalize the proof of \Cref{cor:fermat-catalan}.

FLT for polynomials (\Cref{cor:fermat}) then immediately follows by considering the case when $p = q = r = n \geq 3$ and $u = v = 1$, $w = -1$.

\begin{minted}{lean}
theorem Polynomial.flt {n : ℕ} (hn : 3 ≤ n) (chn : ¬ringChar k | n)
  {a b c : k[X]} (ha : a ≠ 0) (hb : b ≠ 0) (hc : c ≠ 0) 
  (hab : IsCoprime a b) (heq : a ^ n + b ^ n = c ^ n) :
    a.natDegree = 0 ∧ b.natDegree = 0 ∧ c.natDegree = 0
\end{minted}

%% file: 62elliptic.tex
\subsection{Non-parametrizability of $y^2 = x^3 + 1$ (\Cref{cor:elliptic})}
\label{subsec:pf-elliptic}

\subsubsection{Mathematical proof}

As a corollary of \Cref{cor:fermat-catalan}, we can show that $y^2 = x^3 + 1$ is not parametrizable by rational functions of $t$, similarly as in \cite[Proposition 2.3.1]{franz}.

Assume that a parametrization exists, so that $x = m / M$ and $y = n / N$ for some $m, n, M, N \in k[t]$ with $(m, M) = 1$ and $(n, N) = 1$.
By clearing denominators, we obtain $n ^ 2 M ^ 3 = (m ^ 3 + M ^ 3) N ^ 2$. 
From this, one can show that $N^2$ and $M^3$ divide each other.
Using the unique factorization of $N^2$ and $M^3$, we can find $\alpha, \beta \in k^\times$ and $e \in k[t]$ such that $M = \alpha e^2$ and $N = \beta e^3$.
Now the equation reduces to $\beta^2 m^3 + \alpha^3 \beta^2 e^6 = \alpha^3 n^2$, which is a nontrivial solution for the Fermat--Catalan equation with $(p, q, r) = (3, 6, 2)$.
This is a contradiction as the characteristic of $k$ is not $2$ or $3$.

\subsubsection{Formalization}

The statement can be formalized as follows.

\begin{minted}{lean}
def IsConst (x : RatFunc k) := ∃ c : k, x = RatFunc.C c

theorem no_parametrization_y2_x3_1
  (chk : ¬ringChar k | 6) {x y : RatFunc k} (eqn : y ^ 2 = x ^ 3 + 1) : 
    IsConst x ∧ IsConst y
\end{minted}

The formalization is straightforward, but requires a large body of code for algebraic manipulation.
Also, we had to formalize certain number-theoretic properties following from the fact that $k[t]$ is a unique factorization domain (UFD).
For example, that $M^2$ and $N^3$ divides each other 
impling the existence of $c$ such that $M$ and $N$ are associated to $c^3$ and $c^2$ respectively, which is true for any UFD.

\begin{minted}{lean}
theorem associated_pow_pow_coprime_iff
  {a b : k[X]} (ha : a ≠ 0) (hb : b ≠ 0)
  {m n : ℕ} (hm : m ≠ 0) (hn : n ≠ 0)
  (h : Associated (a ^ m) (b ^ n)) (hcp : m.Coprime n) :
    ∃ c : k[X], c ≠ 0 ∧ Associated a (c ^ n) ∧ Associated b (c ^ m)
\end{minted}

%% file: 63davenport.tex
\subsection{Davenport's Theorem (\Cref{cor:davenport})}
\label{subsec:pf-davenport}

\subsubsection{A Non-coprime Variant of Mason--Stothers Theorem}

Davenport's theorem also almost directly follows from Mason--Stothers theorem.
We start with a variant of Mason--Stothers theorem by Stothers \cite{stothers81} that does not require coprimality of $a, b, c$.

\begin{theorem}
\label{thm:mason-stothers-nocoprime}
Let $k$ be any field of characteristic zero, and $a, b, c \in k[t]$ be non-zero polynomials satisfying $a + b + c = 0$.
Then we either have $a, b, c \in k$ or
\[
\max\{\deg(a), \deg(b), \deg(c)\} < \deg (\rad(a)) + \deg (\rad(b)) + \deg (c).
\]
\end{theorem}

Note that we need $k$ to have characteristic zero in \Cref{thm:mason-stothers-nocoprime}. If $\textrm{char } k = p > 0$, then a counterexample is $(a, b, c) = (t^{p+1}, -t(1 + t)^p, t)$.

\begin{proof}
Let $d$ be the greatest common divisor of $a$ and $b$.
Then $a = a_0 d, b = b_0 d, c = c_0 d$ for $a_0, b_0, c_0 \in k[t]$ with $a_0 + b_0 + c_0 = 0$.
Moreover, $\gcd(a_0, b_0) = 1$ so $a_0, b_0, c_0$ are nonzero and pairwise coprime. By applying \Cref{thm:Mason--Stothers} to $(a_0, b_0, c_0)$, we either have $a_0' = b_0' = c_0' = 0$ or
\begin{equation}
\label{eq:ineq}
\begin{gathered}
\max\{\deg(a_0), \deg(b_0), \deg(c_0)\} <  \\ 
\deg (\rad(a_0)) + \deg (\rad(b_0)) + \deg (\rad(c_0)).
\end{gathered}
\end{equation}

Consider the case $a_0' = b_0' = c_0' = 0$. Since $k$ has characteristic zero, $a_0, b_0, c_0 \in k$.
If $d \in k$, then the proof is done. Otherwise, $\deg (d) \geq 1$ by \Cref{lem:nzdeg}, so
\begin{align*}
\max\{\deg(a), \deg(b), \deg(c)\} = \deg(d) & < \deg (\rad(d)) + \deg (\rad(d)) + \deg (d) \\
& = \deg (\rad(a)) + \deg (\rad(b)) + \deg (c)
\end{align*}
and the proof is done too.

Now consider the case where (\ref{eq:ineq}) is true. Then
\begin{align*}
\max\{\deg(a), \deg(b), & \deg(c)\} = \max\{\deg(a_0), \deg(b_0), \deg(c_0)\} + \deg(d) \\
& < \deg (\rad(a_0)) + \deg (\rad(b_0)) + \deg (\rad(c_0)) + \deg (d) \\
& \leq \deg (\rad(a)) + \deg (\rad(b)) + \deg (c_0) + \deg(d) \\
& = \deg (\rad(a)) + \deg (\rad(b)) + \deg (c)
\end{align*}
completing the proof of \Cref{thm:mason-stothers-nocoprime}.
\end{proof}

The variant \Cref{thm:mason-stothers-nocoprime} is formalized as follows.
\begin{minted}{lean}
theorem Polynomial.abc'_char0 [CharZero k]
    {a b c : k[X]} (ha : a ≠ 0) (hb : b ≠ 0) (hc : c ≠ 0)
    (hsum : a + b + c = 0) :
    (a.natDegree = 0 ∧ b.natDegree = 0 ∧ c.natDegree = 0) ∨
      Nat.max₃ a.natDegree b.natDegree c.natDegree + 1 ≤
      (radical a).natDegree + (radical b).natDegree + c.natDegree
\end{minted}

\subsubsection{Mathematical proof}

We now prove Davenport's theorem (\Cref{cor:davenport}), mainly following the proof in Stothers' paper \cite{stothers81}.\footnote{Our proof is slightly more streamlined; we do not divide the proof into cases on whether $\deg(f^3) = \deg(g^2)$ or not.} For non-constant polynomials $f, g \in k[t]$ with $f^3 - g^2 \ne 0$, apply \Cref{thm:mason-stothers-nocoprime} to the zero-sum triple $(-f^3, g^2, f^3 - g^2)$.
The equality case
$$(f^3)' = (g^2)' = (f^3 - g^2)' = 0$$ cannot happen since it would imply $3f^2f' = 0 = 2gg'$ and thus $3 = 0 = 2$.

So we get the inequality
\begin{align*}
    \max \{3 \deg (f), 2 \deg (g) \} & \leq \max \{ \deg(-f^3), \deg(g^2), \deg(f^3 - g^2) \} \\
    &< \deg(\rad(-f^3)) +  \deg(\rad(g^2)) + \deg(f^3 - g^2) \\
    &\le \deg (f) + \deg(g) + \deg(f^3 - g^2).
\end{align*}
This gives two inequalities
\begin{align*}
    3 \deg(f) + 1 & \leq \deg(f) + \deg(g) + \deg(f^3 - g^2) \\
    2 \deg(g) + 1 & \leq \deg(f) + \deg(g) + \deg(f^3 - g^2)
\end{align*}
and adding these two inequalities and rearranging gives the desired inequality.

\subsubsection{Formalization}

The statement of Davenport's theorem (Corollary \ref{cor:davenport}) can be formalized as follows.
\begin{minted}{lean}
theorem Polynomial.davenport [CharZero k]
    {a b : k[X]} (ha : a.natDegree > 0) (hb : b.natDegree > 0)
    (hnz : a ^ 3 - b ^ 2 ≠ 0) :
    a.natDegree + 2 ≤ 2 * (a ^ 3 - b ^ 2).natDegree
\end{minted}

We also formalized a variant of Davenport's theorem that allows arbitrary characteristics, with the cost of assuming coprimality of two polynomials and assuming non-vanishing derivative instead of non-constantness.
Note that we cannot remove all of these assumptions; $k = \mathbb{F}_{2}$ with $(a, b) = (t^4, t^6 + t)$ gives a counterexample.

\begin{minted}{lean}
theorem Polynomial.davenport' {a b : k[X]} (hab : IsCoprime a b) (haderiv : derivative a ≠ 0) (hbderiv : derivative b ≠ 0) :
    a.natDegree + 2 ≤ 2 * (a ^ 3 - b ^ 2).natDegree
\end{minted}

%% file: 7comparison.tex
\section{Comparison with Previous Works}
\label{sec:comparison}

We compare our work to other formalizations of the Mason--Stothers Theorem by Eberl in Isabelle \cite{eberl17} and by Wagemaker in Lean 3 \cite{wagemaker-github,wagemaker18}.\footnote{Note that an unpublished Coq formalization by Assia Mahboubi is also reported in \cite[Chapter 5]{wagemaker18}. We do not compare our work to this as it is not publicly available.}
All three formalizations, including ours, are based on the same proof by Lemmermeyer's note \cite{franz} on the elementary proof of Snyder \cite{snyder00}.
Unlike Snyder's original proof \cite{snyder00} which assumes that $k$ is algebraically closed, all formalizations work
with any field $k$ using radicals, following \cite[Theorem 2.1.4, Corollary 2.1.5]{franz}.

\begin{center}
\begin{table}[h]
    \caption{Comparison of definitions and theorems in different formalizations of Mason--Stothers.}
\resizebox{\columnwidth}{!}{
    \begin{tabular}{c|c|c|c}
        \toprule
         & Eberl \cite{eberl17} & Wagemaker \cite{wagemaker-github} & Ours \\
        \midrule
        \midrule
        Radical & \texttt{radical} & \texttt{rad} & \texttt{radical} \\ 
        \midrule
        $(a, b) = 1\Rightarrow\rad(ab) = \rad(a)\rad(b)$ & \texttt{radical\_mult\_coprime} & \texttt{rad\_mul\_eq\_rad\_mul\_rad\_of\_coprime} & \texttt{radical\_hMul} \\
        \midrule
        $\deg W(a,b) < \deg(a) + \deg(b)$ & \texttt{degree\_pderiv\_mult\_less}\tablefootnote{Eberl formalized $\deg(a'b) < \deg(a) + \deg(b)$ instead and used it twice.} & \texttt{degree\_wron\_le} & \texttt{natDegree\_lt\_add} \\
        \midrule
        $\frac{a}{\mathrm{rad}(a)} | a'$ (\cite[Lemma 2.1.2]{franz}) & \texttt{poly\_div\_radical\_dvd\_pderiv} & \texttt{Mason\_Stothers\_lemma}\tablefootnote{Doest not exactly prove $\frac{a}{\rad(a)} | a'$; see \Cref{subsec:comp-wagemaker} for details.} & \texttt{divRadical\_dvd\_derivative} \\
        \midrule
        \multirow{2}{*}{Mason--Stothers} & \texttt{Mason\_Stothers}  & \multirow{2}{*}{\texttt{Mason\_Stothers}} & \multirow{2}{*}{\texttt{Polynomial.abc}} \\
        & \texttt{Mason\_Stothers\_char\_0} & & \\
        \midrule
        \multirow{2}{*}{Polynomial FLT} & \texttt{fermat\_poly}  & \multirow{2}{*}{-} & \multirow{2}{*}{\texttt{Polynomial.flt}} \\
        & \texttt{fermat\_poly\_char\_0} & & \\
        \bottomrule
    \end{tabular}}
    \label{tab:common}
\end{table}
\end{center}

\subsection{Eberl's Isabelle formalization}
\label{subsec:comp-eberl}

Eberl formalized both the characteristic zero and positive case of Mason--Stothers theorem in Isabelle \cite{eberl17}, as a part of the \emph{Archive of Formal Proofs} (Isabelle-AFP) mathematics library.
Consequently, their formalization is reusable with other definitions and theorems in Isabelle-AFP.
We compare their formalization to ours as follows.

\begin{enumerate}
\item They define the radical $\rad(a)$ on any \emph{factoral semiring}, which is a commutative ring with unique factorization. We define radical in a slightly more general setting of monoids with unique factorization.  

\item They assume the coprimality condition $\mathrm{gcd}(a, b, c) = 1$\footnote{\texttt{cop: "Gcd \{A, B, C\} = 1"} in Isabelle} in Mason--Stothers, but this is equivalent to pairwise coprimality $\mathrm{gcd}(a, b) = \mathrm{gcd}(b, c) = \mathrm{gcd}(c, a) = 1$ we assume by $a+b+c=0$.

\item Their work also formalizes the polynomial version of FLT for any characteristic.
They proved that, when a triple of nonzero coprime polynomials satisfy $a^n + b^n + c^n = 0$ and at least one of $(a^n)'$, $(b^n)'$, or $(c^n)'$ is nonzero,
% \footnote{\texttt{deg: "}$\exists p \in \{\texttt{A,B,C}\} \texttt{. pderiv} (\texttt{p}^n) = 0$ in Isabelle}
 then $n \le 2$.
In other words, nonzero coprime polynomials $a, b, c$ satisfying the Fermat's equation for $n \ge 3$ should have $(a^n)' = (b^n)' = (c^n)' = 0$.
Our formalization of polynomial FLT (\Cref{cor:fermat}) has a strictly stronger conclusion; either the characteristic of $k$ divides $n$, or $a, b, c \in k$.
% \footnote{Our condition implies the conclusion $(a^n)' = (b^n)' = (c^n)' = 0$ of Eberl's version immediately. On the other hand, let $k$ be of characteristic $p > 0$, let $n$ be any number not divisible by $p$, and let $a = t^p$. Then $(a^n)' = (t^{np})' = 0$ holds, satisfying the conclusion of Eberl's, but observe that $a$ is not a constant.}
This is achieved by the simple infinite descent argument in \Cref{subsubsec:pf-fermat-catalan}.
\end{enumerate}

\subsection{Wagemaker's Lean 3 formalization}
\label{subsec:comp-wagemaker}

Wagemaker formalized the Mason--Stothers theorem in Lean 3, in the early days when the Lean \texttt{mathlib} mathematics library was taking shape \cite{wagemaker-github,wagemaker18}.
Consequently, H\"{o}lzl and Wagemaker built a large body of work (``4/5 of the formalization'' according to Wagemaker \cite{wagemaker18})  that formalizes many fundamental notions such as polynomials,  UFDs, greatest common divisor, and coprimality \cite{wagemaker-github}.
Their work was then incorporated into the current \texttt{mathlib}/\texttt{mathlib} library of Lean 3 and 4.
In particular, the design suggestions \cite{wagemaker18} in Wagemaker's work shapes a lot of fundamental APIs in the current \texttt{mathlib} implementation of UFDs.\footnote{For an example, he observed that the notion of greatest common divisor in a general UFD $R$ should have the type of \emph{quotients} of $R$ modulo associated elements, which is now available as \texttt{Associates} in \texttt{mathlib}.}.

Their project was independent of Lean 3's \texttt{mathlib}, however, as it was incorporated \emph{after} its completion.
In contrast, our work builds on the now-mature \texttt{mathlib} of Lean 4, ensuring reusability with existing definitions.
In regards to the formalization of Mason--Stothers theorem, we compare their work to ours as follows.
\begin{enumerate}
\item They work on fields of characteristic zero only, while our formallization allows arbitrary characteristic.

\item They do not formalize further corollaries of Mason--Stothers such as polynomial FLT.

\item Their work misses the proof that a polynomial ring $R[X]$ over a unique factorization domain $R$ is also a unique factorization domain.\footnote{This is represented as a \texttt{sorry} in \texttt{poly\_over\_UFD.lean} of \cite{wagemaker-github}.}
In contrast, our formalization on Lean 4 is complete, and is based on the current \texttt{mathlib} with the proof that $R[X]$ is UFD (and many more, formalized by others).

\item They do not define $a / \rad(a)$ explicitly but instead use $\gcd(a, a')$ to avoid polynomial division. Then they prove
\[
    \deg(a) \le \deg(\gcd(a, a')) + \deg(\rad(a))
\]
as a lemma \cite[Lemma 2.3.1]{wagemaker18}, instead of \Cref{lem:main} in our work.
\end{enumerate}

\begin{table}[ht]
\begin{center}
    \caption{Comparison of Formalizations of the Mason--Stothers Theorem}
    \begin{tabular}{c|c|c|c|c}
        \toprule
        \multicolumn{2}{c|}{} & Eberl \cite{eberl17} & Wagemaker \cite{wagemaker-github,wagemaker18} & Ours \\
        \midrule
        \midrule
        \multicolumn{2}{c|}{Language} & Isabelle & Lean 3 & Lean 4 \\
        \midrule
        \multicolumn{2}{c|}{Complete} & O & X & O \\
        \midrule
        \multirow{2}{*}{Mason-Stothers} & $\text{char}=0$ & O & O & O\tablefootnote{Includes a non-coprime variant (\Cref{thm:mason-stothers-nocoprime}) by Stothers.} \\
        \cmidrule{2-5}
        & $\text{char}>0$ & O & X & O \\
        \midrule
        \multirow{2}{*}{Poly-FLT} & $\text{char}=0$ & O & X & O \\
        \cmidrule{2-5}
         & $\text{char}>0$ & O & X & O\tablefootnote{Stronger conclusion than Eberl \cite{eberl17} as described in \Cref{subsec:comp-eberl}.} \\
        \midrule
        \multicolumn{2}{c|}{Other corollaries} & X & X & O \\
        \bottomrule
    \end{tabular}
    \label{tab:comparison}
\end{center}
\end{table}

%% file: 8futureworks.tex
\section{Future Works}\label{sec:5}

% \textcolor{blue}{We formalized the Mason-Stothers theorem, which is an analog of ABC conjecture for polynomials, using Lean 3. Although the theorem is already formalized, some are incomplete \cite{wagemaker18} or [the reason that ours are better than Isabelle and Coq formalization] \cite{eberl17}. Not only do we give a complete formalization of the theorem, we even provide formalizations of corollaries of the theorem, including FLT for polynomials, nonparametrizability of an elliptic curve, and Davenport's theorem. Also, our formalization is consistent with the current mathlib library, which makes it easy to use our formalization for other purposes.}

We suggest further directions in formalizing generalizations of Mason--Stothers theorem.
\begin{itemize}[label=\textbullet]
    \item Bayat et al. \cite{bayat05} extends the Mason--Stothers theorem to more than three polynomials, using the Wronskian of more than two polynomials.

    \item In algebraic geometry, it is known that rational maps from the projective line to a curve exist only if the curve has genus 0. This immediately proves both FLT for polynomials and non-parametrizability of any elliptic curves, as the Fermat curve $x^n + y^n = 1$ ($n\geq 3$) and elliptic curve $y^2 = x^3 + ax + b$ have genus $> 0$ \cite{franz}.
    However, the current \texttt{mathlib} does not have enough theorems to prove this result (e.g. Riemann--Hurwitz formula).
    % \textcolor{blue}{This can be proven using Riemann-Hurwitz formula which is not formalized in Lean yet.}
        
    \item Mason--Stothers theorem can be thought as the most basic case of ABC over a function field $k(C)$ of a smooth projective curve $C$, when the curve $C$ is the projective line over $k$.
    Mason \cite{mason83} proved the following more general result:
    \begin{theorem}[Mason]
    Let $k$ be an algebraically closed field and $C$ be a smooth projective curve over $k$.
    Let $a, b \in k(C)$ satisfying $a + b = 1$, and let $S$ be a finite subset of points in $C(k)$ containing all the zeros and poles of $a$ and $b$.
    Then either $a, b \in k^\times$ or 
    \[
         \max\{\deg(a), \deg(b)\} \leq 2g - 2 + |S|.
    \]
    \end{theorem}
    When $C = \mathbb{P}^1$, this reduces to the Mason--Stothers theorem:
    a zero-sum coprime triple $a, b, c$ of polynomials gives $(-a/c) + (-b/c) = 1$, and the above inequality becomes
    % \[
    %     \max \{ \deg(a), \deg(b), \deg(c) \} = \max \left\{\deg \left(-\frac{a}{c}\right), \deg \left( - \frac{b}{c}\right) \right\} \leq -2 + |S| = \deg (\rad(abc)) - 1,
    % \]
    \begin{align*}
        &\max \{ \deg(a), \deg(b), \deg(c) \} \\
        &= \max \left\{\deg \left(-\frac{a}{c}\right), \deg \left( - \frac{b}{c}\right) \right\} \\
        &\leq -2 + |S| = \deg (\rad(abc)) - 1,
    \end{align*}
    where $S = \{\text{zeros of }abc\} \cup \{\infty\}$ and $|S| = \deg (\rad(abc)) + 1$.
    Silverman \cite{silverman84} gives a short proof of the theorem using Riemann--Hurwitz formula.

\end{itemize}

%% file: appendix.tex
\appendix
\section{Porting to \texttt{mathlib}}
\label{sec:porting}

% We are in the process of integrating our formalization of the Mason--Stothers theorem to the \texttt{mathlib} library.
% Table \ref{tab:mathlibpr} lists the pull requests made so far to \texttt{mathlib} as of January 7, 2025.

Our formalization of the Mason--Stothers theorem and the polynomial FLT has been integrated into the \texttt{mathlib} library.
Table \ref{tab:mathlibpr} lists the pull requests made, where \href{https://github.com/leanprover-community/mathlib4/pull/15706}{15706} and \href{https://github.com/leanprover-community/mathlib4/pull/18882}{18882} are the main pull requests that include proofs of these main theorems.

\begin{table}[h]
\begin{center}
    \caption{List of pull requests to \texttt{mathlib}.}
    \resizebox{\columnwidth}{!}{
    \begin{tabular}{c|c|c}
        \toprule
        Topic & PR & Descriptions \\
        \midrule
        \midrule
        \multirow{3}{*}[-.5em]{Wronskian} & \href{https://github.com/leanprover-community/mathlib4/pull/14281}{14281} & Prove $B(a, b) + B(b, c) + B(c, a) = 0$ for alternating bilinear $B$. \\
        \cmidrule{2-3}
        & \href{https://github.com/leanprover-community/mathlib4/pull/14243}{14243} & Define Wronskian of polynomials and prove relevant theorems. \\
        \cmidrule{2-3}
        & \href{https://github.com/leanprover-community/mathlib4/pull/18483}{18483} & For coprime $a, b \in k[x]$, $W(a, b) = 0$ if and only if $a' = b' = 0$.  \\
        \midrule
        \multirow{5}{*}[-1em]{Radical} & \href{https://github.com/leanprover-community/mathlib4/pull/14873}{14873} & Define radical of elements. \\
        \cmidrule{2-3}
        & \href{https://github.com/leanprover-community/mathlib4/pull/15331}{15531} & Prove theorems on radicals of coprime elements. \\
        \cmidrule{2-3}
        & \href{https://github.com/leanprover-community/mathlib4/pull/18449}{18449} & \texttt{divRadical} for Euclidean domain. \\
        \cmidrule{2-3}
        & \href{https://github.com/leanprover-community/mathlib4/pull/18452}{18452} & \texttt{divRadical} for polynomials. \\
        \cmidrule{2-3}
        & \href{https://github.com/leanprover-community/mathlib4/pull/20468}{20468} & $\deg(\rad(a)) \le \deg(a)$. \\
        \midrule
        \multirow{2}{*}[-.2em]{Miscellaneous} & \href{https://github.com/leanprover-community/mathlib4/pull/19133}{19133} & For $x, y \in R$ and $u, v \in R^\times$, $(x, y) = 1$ iff $(ux, vy) = 1$. \\
        \cmidrule{2-3}
         & \href{https://github.com/leanprover-community/mathlib4/pull/20190}{20190} & $(a^n)' = 0$ iff $a' = 0$ when $n \ne 0$ in the coefficient ring. \\
        \midrule
        Mason--Stothers & \href{https://github.com/leanprover-community/mathlib4/pull/15706}{15706} & Proof of Mason--Stothers theorem. \\
        \midrule
        \multirow{2}{*}[-.2em]{Polynomial FLT} & \href{https://github.com/leanprover-community/mathlib4/pull/16060}{16060} & Statement of FLT for semirings, allowing nonzero unit solutions. \\
        \cmidrule{2-3}
        & \href{https://github.com/leanprover-community/mathlib4/pull/18882}{18882} & Proof of the polynomial FLT. \\
        \bottomrule
    \end{tabular}
    }
    \label{tab:mathlibpr}
\end{center}
\end{table}